\documentclass[11pt,a4paper]{article}
\usepackage[square,sort&compress,comma,numbers]{natbib} 
\usepackage{amsmath,amsfonts,graphicx,psfrag,bm}
\usepackage[textsize=small]{todonotes}
\usepackage{algorithm}
\usepackage{algorithmic}
\usepackage{booktabs}
\usepackage{csquotes}
\usepackage{comment}
\usepackage{url}
\usepackage{textcomp}
\usepackage{lscape}
\usepackage{multirow}
\usepackage{mathtools}
\usepackage{xcolor}
\usepackage{ulem}
\usepackage{bibentry}
\usepackage{hyperref} 

\definecolor{nclblue}{rgb}{0, 0.24705882, 0.44705882}                     
\definecolor{nclred}{rgb}{0.77647059,0.04705882,0.18823529}         

\hypersetup{
	colorlinks,                      															
	citecolor=nclred,            															
	filecolor=black,               														  
	linkcolor=black,           															
	urlcolor=nclblue                														     
}

\usepackage{ltxtable}
\usepackage{caption}

\defcitealias{national2019traffic}{NCSA, 2021}

\makeatletter
\newcommand\notsotiny{\@setfontsize\notsotiny\@vipt\@viipt}
\makeatother

\setcitestyle{authoryear,open={(},close={)}}

\title{Using Extreme Value Theory to Evaluate the Leading Pedestrian Interval Road Safety Intervention}
\author{Nicola Hewett$^a$, Lee Fawcett$^a$, Andrew Golightly$^b$ \&  Neil Thorpe$^c$}
\date{$^a$School of Mathematics, Statistics and Physics, Newcastle University,\\
  Newcastle upon Tyne, NE1 7RU, UK\\
  $^b$Department of Mathematical Sciences, Durham University,\\
Durham, DH1 3LE, UK\\
  $^c$Jacobs, Rotterdam House, 116 Quayside, \\
  Newcastle Upon Tyne, NE1 3DY, UK}

\begin{document}
\maketitle
\begin{abstract}
Improving road safety is hugely important with the number of deaths on the world’s roads remaining unacceptably high; an estimated 1.35 million people die each year as a result of road traffic collisions (WHO, 2020). Current practice for treating collision hotspots is almost always reactive: once a threshold level of collisions has been overtopped during some pre-determined observation period, treatment is applied (e.g. road safety cameras). Traffic collisions are rare, so prolonged observation periods are necessary.  However, traffic \textit{conflicts} are more frequent and are a margin of the social cost; hence, traffic conflict before/after studies can be conducted over shorter time periods. We investigate the effect of implementing the leading pedestrian interval (LPI) treatment (Van Houten et al. 2000) at signalised intersections as a safety intervention in a city in north America. Pedestrian-vehicle traffic conflict data were collected from treatment and control sites during the before and after periods. We implement a before/after study on post-encroachment times (PETs) where small PET values denote a `near-miss'. Hence, extreme value theory is employed to model extremes of our PET processes, with adjustments to the usual modelling framework to account for temporal dependence and treatment effects.
 
\end{abstract}

\noindent\textbf{Keywords:} Extreme value theory (EVT); traffic conflicts; leading pedestrian interval (LPI); post-encroachment time (PET); before-after analysis; Bivariate threshold excess model 

\section{Introduction}
\label{sec:intro}
Around 1.3 million people are killed every year as a result of road traffic accidents and between 20 to 50 million people suffer from non-fatal injuries -- around half of these are vulnerable road users such as pedestrians \citep{WHO}. The National Highway Traffic Safety Administration found that, in the USA, the second most common cause for injury-classed collisions is left-turning vehicles, and 22\% of crashes at intersections involve vehicles turning left \citepalias{national2019traffic}. Furthermore, 23\% of pedestrian crashes occurred at intersections.\\

With the safety of pedestrians in mind, safety treatments at signalised intersections have been investigated. \textit{``The leading pedestrian interval (LPI) is one treatment that has been implemented at signalised intersections to permit pedestrians to begin crossing several seconds before the release of conflicting vehicle movements"} \citep{van2000field}. This, theoretically, should reduce potential conflicts between pedestrians and vehicles. A significant reduction in potential conflicts will likely lead to a significant reduction in actual collisions. With such an intervention, there is the added benefit of treating potential road safety hotspots \textit{proactively}. Standard road safety interventions are usually analysed \textit{reactively}, once a threshold collision count has been over-topped during some pre-determined observation period. 
 Collisions are rare events and so prolonged observation periods are necessary, with much waiting for collisions to happen to evaluate a treatment using a standard before/after (BA) analysis \citep[perhaps using Empirical Bayes methodology -- see, for example,][]{ hauer1980bias,fawcett2013mobile,fawcett2017novel}.  Working with near-misses, as the LPI intervention does, means not having to wait for collisions to happen; the LPI is implemented by adjusting the signal-phasing and pedestrian interval to provide a walk display of several seconds before the adjacent vehicle green display, making this an efficient and low-cost safety measure. \\
 
BA studies are prominent in road safety intervention evaluations.  Traditionally, road traffic accident (RTA) BA studies are focused on the reduction in frequency and/or severity of collisions from before the intervention, to after \citep[see, for example,][]{hauer1997observational, elvik2002importance, sayed2016evaluating}.
Traffic conflicts are generally defined as a situation in which a pedestrian and a vehicle approach each other in time and space to such an extent that they will collide if their movements remain unchanged. Traffic conflicts are more frequent than collisions, easy to observe and are of marginal social cost \citep{tarko2018surrogate}. As such, traffic conflict BA studies can be conducted over shorter time periods. Furthermore, new technology has been adopted, such as automated video techniques \citep{saunier2007automated} and traffic simulation \citep{wang2018combined}, to allow for detecting and tracking moving objects based on their trajectories; conflict data are thus easily extracted. \\

Commonly, statistical inference is based on averages obtained from datasets, with procedures utilising the central limit theorem being employed -- supporting the use of the standard Normal distribution or associated $t$ distribution.  As we shall discuss in the next section the LPI intervention, and associated traffic conflict data on collision near-misses, justifies the use of the \textit{extremal types theorem} instead.  The associated extreme value distributions, introduced by \citet{fisher1928limiting}, arise as limits for the distribution of maxima (or minima) in sequences of independent, identically distributed random variables.  Typical applications occur in the environmental sciences to model, for example, extreme precipitation events or extreme wind speeds \citep[see, for example,][]{fawcett2006hierarchical, arun2023leading}; for interested readers, we point to the classical reference of \citet{gumbel1958statistics} and, more recently, the tutorial-style text of \citet{coles2013introduction}. In this paper, we focus on the lower tail of the data where small time values indicate a dangerous situation: a near-miss or collision between a pedestrian and left-turning vehicle. \\

The remainder of this paper is organised as follows. A brief description of the data is given in Section~\ref{sec:Data}. The methodology of extreme value theory is explained in Section~\ref{sec:Method} including describing two methods to handle dependence, and the inclusion of covariates. In Section~\ref{sec:Application} we outline the details of the Bayesian inference scheme, before considering the real data application in Section \ref{sec:results}. Conclusions are drawn in Section~\ref{sec:disc}.

\section{Data}
\label{sec:Data}
We have data from a national collision database, investigating the implementation a 5-second LPI at eight intersections in a north American city,
to give pedestrians more time to cross before a left-turning car is released. At each intersection, one crosswalk has been treated. Data were collected from March 1st--June 30th 2018 (before period), and August 1st--October 31st 2018 (after period), with data coming from the same 12-hour period each day (8:00--20:00) to mitigate potential confounders such as cyclic variation.  Data spanning exactly the same time period are also available for a further seven intersections that have not been treated with the LPI intervention, for comparison purposes.  \\

\begin{figure}[h!]
\centering
\includegraphics[scale=0.7]{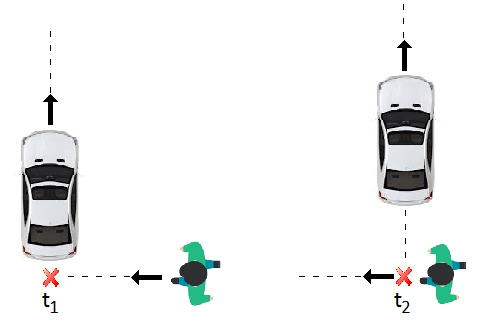}
\caption{Post-encroachment times (PET), $PET=t_2-t_1$.}
\label{fig:T1T2diagram}
\end{figure}

In this dataset, a conflict between a vehicle and a pedestrian is indicated using \textit{post-encroachment time} (PET). PET is the time between the moment the first road user passes the conflicting point, $t_1$, and the moment the second user reaches that point, $t_2$. The positions of the vehicle and pedestrian are shown in Figure \ref{fig:T1T2diagram}, where $\text{PET} = t_1 – t_2$.  We have the minimum PET in 10-minute intervals over the 12-hour study period in each day, for all intersections.  

PETs $<$ 15 seconds were recorded. As $\text{PET} = t_1 - t_2$, if $t_1 = t_2$ then we have a collision between pedestrian and vehicle. Small PET values imply a near-miss, a value close to zero implying a dangerous situation. Our aim is to model extremely small PET values using extreme value theory (EVT), and through this modelling template investigate differences between extremes from periods before and after the LPI treatment was introduced. \\

\begin{figure}[h!]
     \centering
     \includegraphics[]{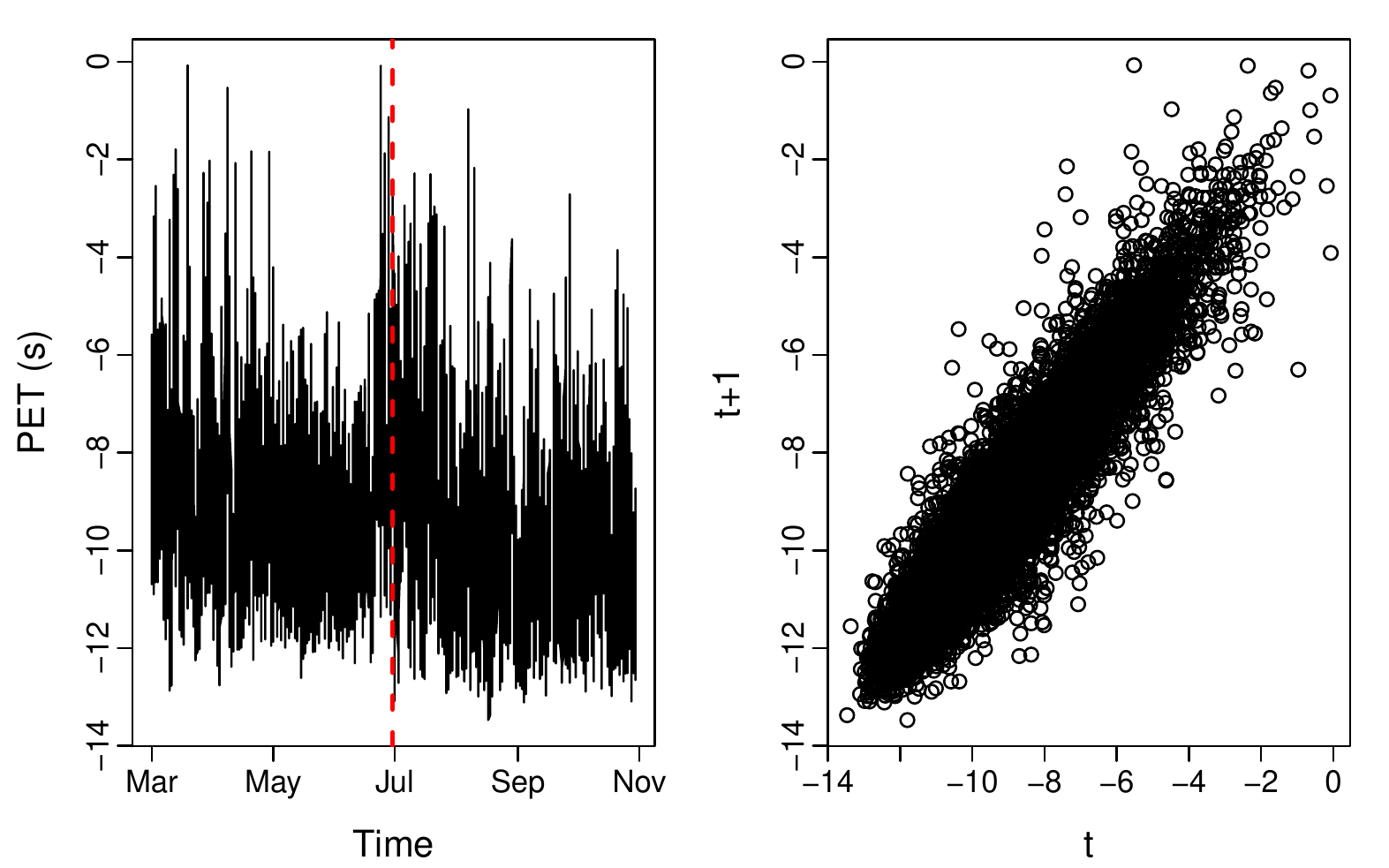}
     \caption{Left: Time series plot of negated PET values (seconds) before and after intervention at intersection 1, the vertical red line showing the start of the after period. Right: Temporal dependence of observations at neighbouring time points for intersection 1 ($t$ on $x$-axis against $t+1$ on $y$-axis).}
     \label{fig:TimeTime}
 \end{figure}

In order to use standard methods from the EVT toolkit directly (designed for analysing `large' extremes), we negate our series of PET values at each location, thus switching the focus from very small values to very large values to identify dangerous situations in our series. Figure \ref{fig:TimeTime} shows a time series plot of these negated PET values, and a plot of observations at neighbouring time points, for one of the intersections at which the LPI was applied. Note the apparent systematic decrease in PET values after the LPI implementation at this specific intersection; note also the clear temporal dependence between consecutive data values, persisting into the extremes of the process.


\section{Methodology}
\label{sec:Method}

\subsection{Background}
EVT has become popular in traffic conflict analyses.
\citet{zheng2019bayesian} use a peaks over threshold approach including covariates in the scale parameter for crash estimation; furthermore, they also use EVT on block maxima from a traffic conflict BA study \citep{zheng2019full}. \citet{wang2019crash} use bivariate EVT to predict annual crash frequencies at intersections. \citet{fu2021multi} use hierarchical EVT modelling on traffic conflict extremes for crash estimation. \citet{guo2020hierarchical} analysed the effectiveness of LPI treatment at two signalised intersections in Vancouver, Canada. They modelled the scale and shape parameters of the GPD as a function of a treatment indicator, a period indicator (before/after), and an interaction of the two variables where dependence between consecutive extremes was removed through a declustering approach. In this paper we use EVT in an attempt to capture treatment effects from the LPI intervention described in Section~\ref{sec:intro}. Our method aims to maximise data usage by performing a threshold-based analysis \citep[as opposed to a block maxima analysis; see, for example,][]{coles2013introduction}. We also allow for a dependence between consecutive extremes, as identified in Figure~\ref{fig:TimeTime}, through a first-order Markov chain structure; and a treatment effect through linear modelling of one of the parameters in the extreme value model used.  \\

EVT provides a framework for estimating the probability of extreme events. The extremal types theorem gives limiting results for the extremes of our process, providing a range of techniques for modelling the tail behaviour of our random variable without any assumptions about the underlying distribution of the data itself. 
 There are two main approaches: the so-called ``block maxima approach'', and the ``threshold-based approach''. In the former, a series of observations are blocked into fixed time intervals (e.g. years, giving rise to an ``annual maxima analysis'') and the maximum observation of each block is extracted to create our series of extremes. In a threshold-based analysis, all observations exceeding a suitable, pre-defined threshold are classified as extreme.  This study proposes the use of a threshold-based approach on the premise of including more data in the analysis than the block maxima approach, which can be wasteful through discarding all but the most extreme observations in each block.  It is hoped that the inclusion of more datapoints will lend greater precision to the analysis, including reduced uncertainty in our estimated treatment effect.  

\subsection{Classifying and modelling extremes}
 Let {$X_n$} denote a stationary sequence of random variables with common distribution function $F$, and let $M_n =$ max$\{X_1 , \ldots , X_n \}$.
It is typically the case that, as $n \to \infty$,
\begin{equation}
  Pr(M_n \leq x) \approx F^{n\theta}(x), 
  \label{eq:Fn}
\end{equation}
where $\theta \in (0, 1)$ is known as the extremal index; see, for example, \citet{leadbetter1988extremal}. As $\theta \to 0$ there is increasing dependence in the extremes of the process; for an independent process, $\theta = 1$, as $Pr(M_n \leq x) = Pr(X_1, X_2, \ldots, X_n \leq x) = Pr(X_1\leq x) \times Pr(X_2\leq x) \times \ldots=F^n(x)$. Initially concerned with the independent case (i.e. $\theta =1$), classical EVT sought families of limiting models for $F^n$, without reference to the marginal distribution function $F$, as any small discrepancies in $F$ could lead to large discrepancies in $F^n$. \\

Examining the behavior of $M_n$ as $n \to \infty$ gives rise to the Extremal Types Theorem \citep[see][]{fisher1928limiting, gnedenko1943distribution}. The limiting distribution of $M_n$ is degenerate, that is, the distribution converges to a single point on the real line with probability 1, this single point being the upper endpoint of $F$. This is analogous to the sample mean $\bar{X}$ converging to the population mean $\mu$ with certainty in the Central Limit Theorem. The Extremal Types Theorem states that if there exists sequences of constants $a_n > 0$ and $b_n$ such that, as $n \to \infty$,
\begin{equation*}
    Pr\left\{(M_n - b_n)/a_n \leq x\right\} \to G(x)
\end{equation*}
for some non-degenerate distribution $G$, then $G$ is of the same type as one of the following distributions:

\begin{equation*}
    \begin{aligned}
        I: G(x) &= \exp\{-\exp(-x)\}, \hspace{0.5cm} -\infty<x<\infty;\\
        II: G(x)&=\begin{cases}
          0 & x\leq 0,\\
          \exp(-x^{-\alpha}) & x>0, \alpha>0; \\
        \end{cases}\\
        III: G(x)&= \begin{cases}
          \exp\{-(-x)^\alpha\} & x<0,\alpha>0,\\
          1 & x\geq 0.
        \end{cases}
    \end{aligned}
\end{equation*}
Distributions $I$, $II$ and $III$ have become known as the Gumbel, Fr\'{e}chet and Weibull types (respectively), and are known collectively as the extreme value distributions. For both the Gumbel and Fr\'{e}chet distributions the limiting distribution $G$ is unbounded; that is, the upper endpoint tends to $\infty$. The Weibull distribution has a finite upper bound. It should be noted that the Extremal Types Theorem does not ensure the existence of a non–degenerate limit for $M_n$; nor does it specify which of types $I$, $II$ or $III$ is applicable if a limit distribution does exist (i.e. in which \textit{domain of attraction} the distribution of $G$ lies). However, when such a distribution does exist, we find that, by analogy with the Central Limit Theorem, the limiting distribution of sample maxima follows one of the distributions given by the Extremal Types Theorem, no matter what the parent distribution $F$.

\subsubsection{Block maxima: The Generalised Extreme Value distribution}
\label{sec:GEV}
\citet{von1954distribution} and \citet{jenkinson1955frequency} independently derived a distribution which encompasses all three types of extreme value distribution; the Generalised Extreme Value distribution (GEV).
The GEV is the limiting model for $F^n$, with distribution function (d.f.)
\begin{equation*}
    \mathcal{G}(y;\mu,\sigma,\xi) =\begin{cases} 
    \exp\left[-(1+\xi(y-\mu)/\sigma)^{-1/\xi}\right], & \xi \neq 0\\
    \exp\left[-\exp(-(y-\mu)/\sigma)\right], & \xi=0 
    \end{cases}
\end{equation*}
defined on $\{y: 1+ \xi(y-\mu)/\sigma > 0\}$, where $-\infty<\mu<\infty$, $\sigma>0$ and $-\infty<\xi<\infty$ are parameters of location, scale and shape, respectively.
It can be shown that $F^{n\theta}$ is also GEV with d.f. $\mathcal{G}(y; \mu^*, \sigma^*, \xi)$, provided long-range dependence is negligible; technically, this requires Leadbetter's $D(u_n)$ condition to hold -- see \citet{leadbetter1983extremes}. Hence, for block maxima, in practical terms short-range dependence can be ignored. In practice, the GEV is used to model maxima over some convenient calendar unit -- usually years.  However, choosing a suitable block size can sometimes be problematic.  In our application, in which data span only a period of months, years would not be appropriate; blocks would need to be large enough for the limiting results to hold, yet if they're too large there will be too few maxima from which to make inferences. Additionally, as discussed above, this method  generally can be hugely wasteful of data -- discarding, as it does, all but the block maxima.

\subsubsection{Threshold excesses: The Generalised Pareto Distribution}
\citet{pickands1975statistical} showed that for $\theta = 1$ and $u$ large, ($X - u|X > u$) follows a Generalised Pareto Distribution (GPD) with d.f.
    \begin{equation}
        \mathcal{H}(y)=\begin{cases}
        1-(1+\xi y/\tilde{\sigma})^{1/\xi}, & \xi \neq 0 \\
        1-\exp(-y/\tilde{\sigma}), & \xi=0
        \end{cases}
        \label{eq:GPDeq}
    \end{equation}
defined on $y>0$, with scale and shape parameters $\tilde{\sigma}$ and $\xi$ respectively. Here, $\tilde{\sigma}$ is related to the parameters in the corresponding GEV for block maxima through $\tilde{\sigma} = \sigma + \xi(u - \mu)$.  Threshold methods classify observations as extreme if they exceed some high threshold, usually denoted $u$; then, the GPD in Equation (2) is fitted to the excesses over this threshold.  Graphical diagnostics are available for the selection of a suitable threshold, as will be demonstrated in Section 3.  Unlike the case of modelling block maxima with the GEV, powering $F^n$ by $\theta$, as described by Equation (\ref{eq:Fn}), does not lead to another extreme value distribution whose parameters have absorbed the extremal index; thus, careful consideration of extremal dependence is required (and, unlike consecutive block maxima, extremal dependence is usually present between consecutive threshold excesses).

\subsection{Handling dependence}
\label{sec:Depend}
As discussed, our aim here is to maximise precision by including as much data in the analysis as possible; hence, we will proceed with a threshold-based approach to analysis.  In this Section, we describe two methods for handling temporal dependence present between consecutive threshold excesses: a declustering approach, leading to the commonly-used ``peaks over threshold" analysis of a filtered subset of extremes, and an approach that explicitly models the transition from one extreme to the other through a bivariate extreme value model.  As the second plot in Figure \ref{fig:TimeTime} reveals, even above a high threshold there appears to be dependence between consecutive PET values.  At busier times of the day -- perhaps early morning or late-afternoon -- we might expect more pedestrians to be using the crosswalks at each of our intersections, and a greater number of vehicles turning into the intersections, perhaps resulting in a greater number of near-misses (with an associated clustering of small PET values) at these times.  Clustering of extremes is common in many other applications of the threshold approach to extreme value modelling -- for example, dependence between consecutive temperature or wind speed extremes, and serial correlation in extremes obtained from financial time series.  Ignoring such dependence will likely lead to under-estimated uncertainty measures (for example, confidence intervals that are unrealistically narrow); see, \citet{shi1992joint,barao1999extremal}.  

\subsubsection{Declustering}
\label{sec:declust}
The aim of this approach for handling extremal dependence is to
extract a series of independent threshold excesses, justifying the use of $\theta \approx 1$ in Equation~(\ref{eq:Fn}).  An auxiliary `declustering parameter', say $\kappa$, is chosen and a cluster of threshold excesses is then deemed to have terminated once at least $\kappa$ consecutive observations fall sub-threshold. This is repeated over the entire series to identify clusters of excesses. Then, the maximum (or `peak') observation from each cluster is extracted, and the GPD fitted to the set of (hopefully independent) cluster peak excesses.\\

This approach is often referred to as the ``peaks over threshold" approach \citep[POT,][]{davison1990models} and is the most commonly-used approach for dealing with clustered extremes. Although this approach is quite easy to implement, there are issues surrounding the choice of $\kappa$.  If $\kappa$ is too small, the cluster peaks will not be far enough apart to safely assume independence; if $\kappa$ is too large, there will be too few cluster exceedances on which to form our inference (and of course, the approach will be wasteful of data). Furthermore, parameter estimates have been shown to be sensitive to the choice of $\kappa$ \citep{fawcett2012estimating}. Given how commonplace POT analyses for threshold excesses have become, we will include results based on this approach as a baseline for comparing our results using a first-order extreme value Markov chain for modelling dependence.  

\subsubsection{Modelling dependence: First-order extreme value Markov chain}
To avoid declustering, we can account for dependence between consecutive extremes by assuming a first-order Markov structure. For example, we can assume the following joint density for our series of (negated) PET values at each intersection: 
    \begin{equation}
        \frac{\displaystyle \prod_{i=1}^{n-1}g(y_i,y_{i+1};\Theta)}{\displaystyle \prod_{i=2}^{n-1}g(y_i;\Theta)},
        \label{eq:biLike}
        \end{equation}
where $\Theta$ is a generic parameter vector.  In a threshold excess context, univariate contributions to the denominator in Equation (\ref{eq:biLike}) are given through the GPD (on differentiation of Equation (\ref{eq:GPDeq})). Appealing to bivariate EVT, transformation from GPD to unit Fr\'{e}chet margins \citep[see, for example,][]{coles2013introduction} provides a range of models to use for contributions to the numerator in Equation (\ref{eq:biLike}), the most commonly-used being the logistic family with d.f.:
\begin{equation}
    G(y_{i}, y_{i+1}) = \exp\left\{-\left(y_{i}^{-1/\alpha}+y_{i+1}^{-1/\alpha}\right)^{\alpha}\right\}.
    \label{eq:joint}
\end{equation}
Here, $\alpha \in (0,1)$ controls the extent of extremal dependence in the process, with independence giving $\alpha=1$ and $\alpha \to 0$ corresponding to increasing levels of extremal dependence.  Differentiation of Equation (\ref{eq:joint}), with careful censoring when one or both of $(y_{i},y_{i+1})$ lie sub-threshold, gives pairwise contributions to the numerator in Equation (\ref{eq:biLike}).  Interested readers are referred to \citet{coles2013introduction} for further information and a more detailed discussion of bivariate EVT more generally.   Where direct evaluation of Equation (\ref{eq:Fn}) is necessary -- for example, when obtaining quantiles from the fitted distribution (often used as estimates of \textit{return levels} in applications of EVT) -- \citet{fawcett2012estimating} provide an approximation to the extremal index $\theta$ based on the estimated logistic dependence parameter $\alpha$.  

\subsection{Including covariates}
When the data admits non-stationarity -- for example trend, or a dependence on covariates -- we can attempt to incorporate this non-stationarity through linear modelling of the GEV or GPD parameters.  Generally, we can write the extreme value parameters in the form $h(X^T\beta)$, where $h$ is a specified function, $\beta$ is a vector of parameters and $X$ is the model vector. Recall that the GPD($\tilde{\sigma}, \xi$) arises from the GEV($\mu,\sigma,\xi$), where the GPD scale parameter is a function of the GEV location and shape parameters. Thus, attempting to model any trend in our threshold excesses is usually done through linear modelling of the scale parameter, $\tilde{\sigma}$. The PET data we are investigating have before/after time implications at each of the fifteen intersections; hence, we attempt to capture the treatment effect through the following parameterisation of $\tilde{\sigma}$:
    \begin{equation}
        \tilde{\sigma}_t=\exp(\beta_0 + \beta_1 t),
        \label{eq:beta_1}
    \end{equation}
to respect the positivity of $\tilde{\sigma}$ and where $t=0$ in the before period and $t=1$ in the after period. Hence, at each intersection we might use the slope parameter $\beta_1$ as a proxy for our LPI treatment effect; an estimate of $\beta_1$ that might be deemed significantly different from zero might be indicative of a treatment effect at an intersection.  

\section{Application}
\label{sec:Application}

\subsection{Threshold selection}
The \textit{threshold stability property} of the GPD means that if it is a valid model for excesses over some threshold $u_0$, then it is valid for excesses over all thresholds $u > u_0$. Furthermore, for all $u > u_0,$ E$[X - u|X > u]$ is a linear function of $u$.  In practice, this expectation can be estimated empirically as the sample mean of the excesses over $u$. This leads to the mean residual life plot \citep[MRL plot; see, for example,][]{coles2013introduction}: a graphical procedure for identifying a suitably high threshold for modelling extremes via the GPD in which mean excesses over $u$ are plotted against a range of values for $u$, and the optimal threshold is chosen at the lowest point above which we observe linearity in the plot. MRL plots are a standard diagnostic tool for threshold selection and have been constructed using PET data at each of our intersections to provide site-wise thresholds to identify extremes.  

\subsection{Bayesian inference } 
Section~\ref{sec:Method} covers methodology for modelling extreme values, with a particular focus on threshold methods and handling temporal dependence. We choose the Bayesian paradigm within which to make inferences on these models.  A typical statistical analysis might formulate the likelihood function for the assumed statistical model, maximising this with respect to the parameters in that model to obtain their \textit{maximum likelihood estimates}: values that maximise the likelihood that the process described by the model produced the data that were actually observed.  In a classical sense these are sample-based estimates of fixed but unknown quantities.  In the Bayesian paradigm, the likelihood is merely an ingredient in the inferential process; via Bayes Theorem, it is combined with the density of the \textit{prior distribution} to provide a \textit{posterior distribution} for the parameters of interest -- in effect an update in our beliefs about the parameters after having observed some data, relative to our beliefs before observing these data.  With a careful choice of prior distribution, in some cases it is possible to obtain the precise form of posterior distribution analytically; for example, in the \textit{conjugate} case, both the prior and posterior are from the same family of distributions and the application of Bayes Theorem is trivial. \\

Crucially, the interpretation of the model parameters is different in the Bayesian setting: rather than being fixed (but unknown) constants, as in the classical setting, the parameters are now regarded as random variables, with a distribution (prior and posterior).  This means that in the Bayesian setting, confidence intervals (for example) have a much more natural interpretation, with there being a probability of 0.95 that the parameter falls within the bounds of the 95\% Bayesian confidence (or credible) interval.  But the advantages of working within the Bayesian setting stretch beyond the interpretation of resulting confidence intervals. For example, by their very nature extremes are scarce, and being able to supplement an analysis with more information through the prior distribution has the potential to increase estimation precision.  For example, \citet{fawcett2006hierarchical} record up to an 82\% reduction in uncertainty in estimated wind speed extremes when comparing Bayesian and frequentist approaches to inference. In the analysis of rainfall extremes, \citet{walshaw2003modelling} noted an 84\% reduction in the posterior standard deviation in their estimates when using informative priors formed through discussions with a hydrologist, compared to the corresponding standard errors from a maximum likelihood analysis.  Also, when working with extreme value models, we do not need to worry about the regularity conditions surrounding maximum likelihood estimation of the shape parameter $\xi$, resulting in maximum likelihood estimates being unobtainable when $\xi<-0.5$ -- for full details, see \citet{coles2013introduction}.   \\

Bayesian inference is commonplace in road safety BA studies. A popular method is \textit{empirical Bayes}, which is used to account for regression to the mean effects \citep[see, for example:][]{hauer1980bias,fawcett2013mobile}. More recently, \textit{full Bayes} methods have been used to filter effects of regression to the mean from genuine treatment effects in road safety schemes \citep[see, for example:][]{el2012measuring, heydari2014speed}.  As \cite{fawcett2013mobile} discuss, a full Bayes analysis can provide a more flexible inferential framework with a range of prior distributions beyond the conjugate case being available; it can also provide a more realistic assessment of uncertainty in estimated treatment effects.  

\subsection{Prior specification}

In the absence of any expert prior information regarding our model parameters, and for ease of computation, we adopt fairly uninformative, independent prior distributions.  Recall that, for each intersection, we adopt a bivariate threshold excess model for negated PET values exceeding a threshold (identified and validated through use of an MRL plot).  The margins are assumed GPD with a linear model enabling the scale parameter to vary between before and after periods via Equation~(\ref{eq:beta_1}); for the dependence between successive threshold excesses we adopt a logistic model with a parameter quantifying the degree of serial correlation present.  Thus, our parameter vector at each intersection can be written as:
\[
\Theta = (\beta_{0}, \beta_{1}, \xi, \alpha)^{T},
\]
for which we set the following prior:
\[
\pi(\Theta) = \pi(\beta_0)\pi(\beta_1)\pi(\xi)\pi(\alpha),
\]
and where
\[
\beta_{0}\sim N(0, 10), \quad \beta_{1}\sim N(0, 10), \quad \xi \sim N(0, 100) \quad \text{and} \quad \alpha \sim U(0,1).
\]

\subsection{MCMC sampling}

It is often the case that the posterior distribution cannot be found analytically. For example, in the case of the GPD, there exists no conjugate prior specification for the model parameters, and so we cannot easily find the posterior distribution $\pi(\Theta | \boldsymbol{x})$; we cannot plot this distribution, nor find its moments etc.  Fortunately, techniques to sample from the posterior have been developed and are now routinely used, thanks to their inclusion in many popular statistical software packages. The most common approach is to use a \textit{Markov chain Monte Carlo} (MCMC) scheme \citep[see, for example,][]{casella1992explaining}.  A simple MCMC algorithm is a Normal random walk Metropolis-Hastings (MH) scheme \citep{hastings1970monte}.  Although full details are omitted here, it is noted that -- according to some careful `tuning' of the algorithm through the choice of variance in the Normal random walk updating for the parameter vector $\Theta$ -- it is possible to optimise the algorithm, with the sample returned forming a Markov chain whose stationary distribution is the posterior distribution $\pi(\Theta|\boldsymbol{x})$.  Depending on the starting values chosen, an initial period $B$ might be discarded as `burn-in', to ensure the sample used is indeed from the stationary distribution; to minimise the effects of autocorrelation, it is also common to `thin' the sample and use posterior draws from just every $k$th iteration.  \\

In our MCMC scheme we set initial values for all parameters to their prior means, using a simple MH random walk update to give successive draws
\begin{equation*}
    (\beta_0^{(j)}, \beta_1^{(j)}, \xi^{(j)}, \alpha^{(j)}), \hspace{1cm} j=1,\ldots,10^5,
\end{equation*}
after thinning by every $k=10$ iterations to obtain sufficiently low autocorrelation between realisations and removing the first $B=2000$ iterations as burn-in. A pilot run was implemented to help tune the scheme, with methods from \cite{roberts2001optimal} being used to quickly optimise the algorithm.

\subsection{Results}
\label{sec:results}
\begin{figure}[t]
\centering
\includegraphics[width=\textwidth]{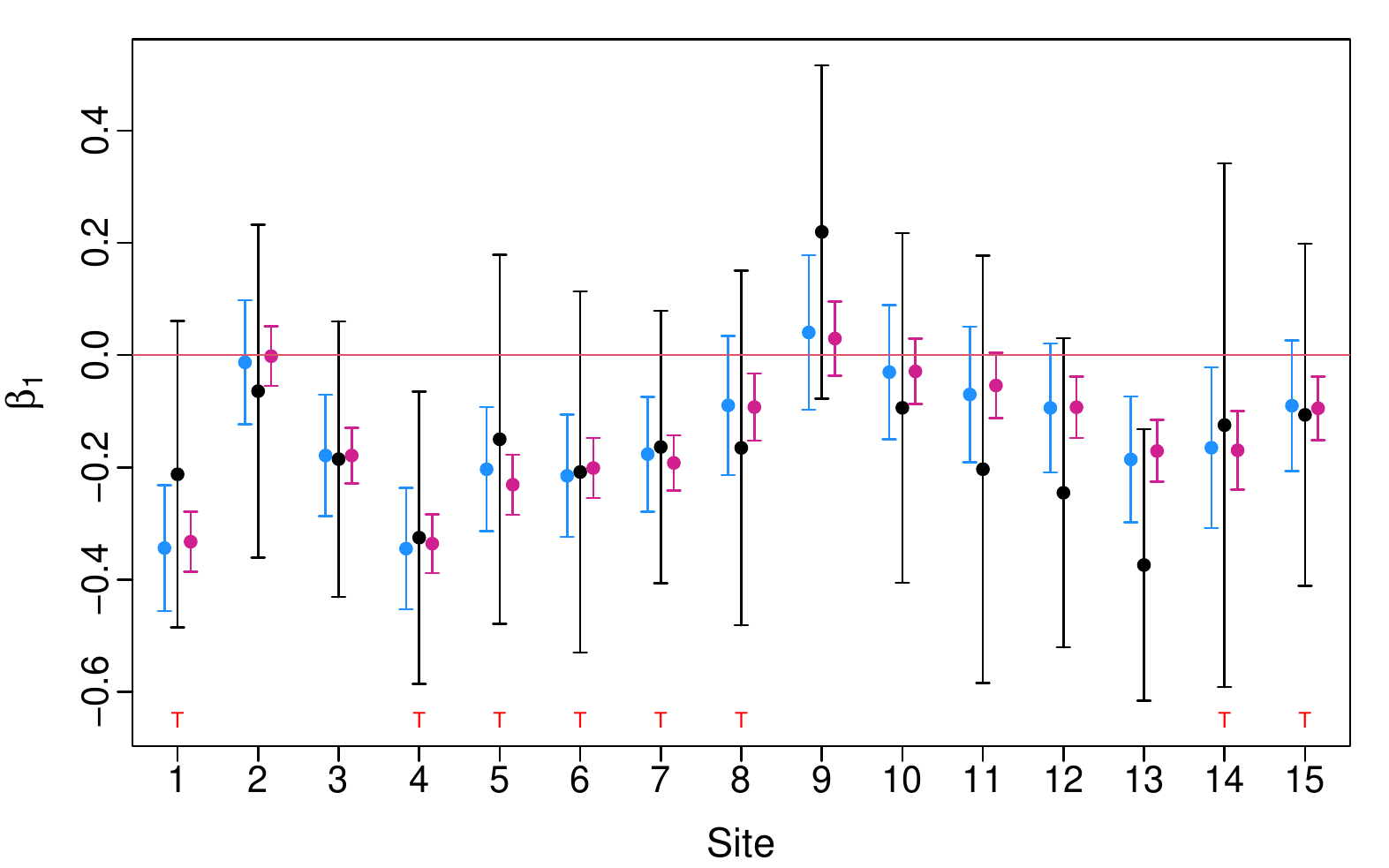}
\caption{Posterior means and 95\% CIs for $\beta_1$ over all sites from ignoring dependence (pink), declustering (grey) and the logistic model (blue). Treated sites denoted with ``T" above the $x$-axis.} 
\label{fig:AllMethod2}
\end{figure}

Figure \ref{fig:AllMethod2} shows the posterior means and 95\% credible intervals for $\beta_1$ over all sites, from analyses that (i) ignore dependence, (ii) filter out dependence through declustering, and (iii) explicitly model the dependence via our first-order extreme value Markov chain model. The treated sites are denoted with a ``T" on the $x$-axis. In our model, $\beta_1$ captures the treatment effect as the slope term in our linear predictor for the GPD scale parameter. \\

When we ignore dependence we use all threshold excesses and hence our credible intervals are relatively narrow owing to the maximal use of extreme data. However, we are violating the assumption that consecutive threshold excesses are independent; as such, these credible intervals are likely to be unrealistically narrow. Declustering (here with $\kappa=10$) removes this dependence, but at a cost: reduced datasets, with site 1 (for example) now having just 107 threshold excesses post-declustering (from an original 17,467 raw excesses). We have done nothing to `optimise' the declustering interval $\kappa$ here, and it could be that our choice of $\kappa$ is unnecessarily large resulting in a procedure that is wasteful of data (giving relatively wide credible intervals). However, even if we were to investigate an optimal choice of declustering interval $\kappa$, our extreme value Markov chain model for explicitly modelling dependence is probably a superior approach here, as it maximises data usage whilst also making some effort to capture the dependence in the series of threshold excesses.  \\

 As $\beta_1$ corresponds to the treatment effect, a value of $\beta_1 <0$ shows a successful treatment as this indicates smaller (negated) PET values in the after period (in other words, larger values on the raw PET scale, meaning a move away from a near-miss/actual collision). Of course, there is uncertainty in our inference, so we look for 95\% credible intervals that are wholly negative to identify a successful treatment effect.  In the majority of the treated sites $\beta_1 = 0$ lies outside the range of the 95\% credible intervals, hence we conclude these sites have seen an improvement post-treatment, with increased PETs.  Notice that, for some sites, treatment effects have not been identified under the POT approach using declustered extremes; the loss in precision, owing to the use of much smaller datasets, often results in wider credible intervals that include zero.  Examples include sites 1, 5, 6, 7 and 14 -- all sites that have been treated with the LPI intervention, and whose credible intervals for $\beta_1$ are wide enough to include zero under the POT approach (but under our extreme value Markov chain model these intervals are wholly negative).      \\

The numerical summaries of the marginal posterior distributions for each parameter, including their mean and 95\% credible intervals, are presented in Table~\ref{tab:resultsEVT} for sites 1, 2, and 3, along with the chosen threshold. The full results for all sites can be found in the appendix, specifically in Table~\ref{tab:FullResults}. For all sites and analyses, the posterior mean of $\xi$ is negative, meaning the GPD has a ``light tail" and is upper bounded. This implies that the distribution has a low probability of producing extreme events. Similarly, as seen in Figure~\ref{fig:AllMethod2} for the posterior summaries of $\beta_1$, the posterior credible intervals for parameters $\beta_0$ and $\xi$ when using declustering are much wider than those from the logistic model. Furthermore, the declustering analysis results in the posterior mean estimates for $\xi$ being more negative meaning we would expect an even lower probability of extreme events. Hence, including all of the data has hugely improved the parameter estimates in terms of uncertainty and has altered our interpretation on the occurrence of extreme events.  

\begin{table}[ht]
\centering
\resizebox{\textwidth}{!}{%
\begin{tabular}{c|cccccc}
\textbf{Model}                     & \multicolumn{1}{l}{\textbf{Site}} & \multicolumn{1}{l}{Threshold, $u$} & \multicolumn{1}{l}{} & $\beta_0$        & $\beta_1$          & $\xi$               \\ \hline
\multirow{2}{*}{Ignore Dependence} & \multirow{6}{*}{1}                & \multirow{6}{*}{-5.6930}           & Mean                 & 0.4108           & -0.3324            & -0.1809             \\
                                   &                                   &                                    & 95\% CI              & (0.3253, 0.4963) & (-0.3861, -0.2786) & (-0.2360, -0.1259)  \\
\multirow{2}{*}{Declustering}      &                                   &                                    & Mean                 & 0.9363           & -0.2123            & -0.3516             \\
                                   &                                   &                                    & 95\% CI              & (0.5911, 1.2816) & (-0.4853, 0.0607)  & (-0.5916, -0.1115)  \\
\multirow{2}{*}{Logistic}          &                                   &                                    & Mean                 & 0.3981           & -0.3436            & -0.1649             \\
                                   &                                   &                                    & 95\% CI              & (0.3015, 0.4948) & (-0.4559, -0.2314) & (-0.2221, -0.1077)  \\ \cline{2-7} 
\multirow{2}{*}{Ignore Dependence} & \multirow{6}{*}{2}                & \multirow{6}{*}{-4.8800}           & Mean                 & 0.1058           & -0.0019            & -0.1373             \\
                                   &                                   &                                    & 95\% CI              & (0.0147, 0.1968) & (-0.0549, 0.0512)  & (-0.1949,  -0.0797) \\
\multirow{2}{*}{Declustering}      &                                   &                                    & Mean                 & 0.6136           & -0.0642            & -0.2552             \\
                                   &                                   &                                    & 95\% CI              & (0.2878, 0.9394) & (-0.3605, 0.2321)  & (-0.4900, -0.0204)  \\
\multirow{2}{*}{Logistic}          &                                   &                                    & Mean                 & 0.1297           & -0.0129            & -0.1454             \\
                                   &                                   &                                    & 95\% CI              & (0.0251, 0.2344) & (-0.1234, 0.0975) & (-0.2082, -0.0826)  \\ \cline{2-7} 
\multirow{2}{*}{Ignore Dependence} & \multirow{6}{*}{3}                & \multirow{6}{*}{-5.9600}           & Mean                 & 0.4687           & -0.2308            & -0.2119             \\
                                   &                                   &                                    & 95\% CI              & (0.3967, 0.5408) & (-0.2841, -0.1775) & (-0.2538, -0.1700)  \\
\multirow{2}{*}{Declustering}      &                                   &                                    & Mean                 & 0.9325           & -0.1498            & -0.3003             \\
                                   &                                   &                                    & 95\% CI              & (0.6369, 1.2280) & (-0.4786, 0.1790)  & (-0.5286, -0.0719)  \\
\multirow{2}{*}{Logistic}          &                                   &                                    & Mean                 & 0.4695           & -0.2033            & -0.1910             \\
                                   &                                   &                                    & 95\% CI              & (0.3858, 0.5533) & (-0.3139, -0.0926) & (-0.2368, -0.1452) 
\end{tabular}%
}
\caption{Numerical posterior summaries for model parameters at intersections 1, 2 and 3. For each parameter we show the posterior mean and 95\% credible interval from the three different models (ignoring dependence, declustering and logistic) and the chosen threshold. }
\label{tab:resultsEVT}
\end{table}

\section{Discussion}
\label{sec:disc}
An extreme value Markov chain was proposed to conduct a traffic conflict-based BA safety evaluation, modelling threshold excesses with the GPD marginally, and using a bivariate extreme value model for the temporal evolution of our extremes, at each intersection. Our approach combines traffic conflicts at different sites (treatment sites and control sites) and for different periods (the before period and the after period) to estimate potential treatment effects of the LPI intervention. 
 Pedestrian traffic conflict data were collected from the treatment and control sites during the before and after periods using automated computer vision analysis techniques. The treatment effects were measured through linear modelling in the scale parameter of the GPD; we avoided unnecessary data wastage, common in the POT approach to analysis, by accounting for temporal dependence and thus including all threshold excesses in the analysis.  \\
 
Use of the bivariate logistic model to account for extremal dependence in our PET processes resulted in narrower credible intervals for our defined treatment effect than a declustering approach, and was also more successful in correctly identifying sites that had been treated.  The posterior distributions for our treatment effect parameter coincide for the model ignoring dependence and our extreme value Markov chain model, with similar posterior means. This is also true for the other parameters in our model. However, although ignoring dependence and fitting a GPD to all threshold excesses results in the narrowest credible intervals, when we ignore dependence our model is no longer valid as our processes clearly exhibit dependence, even in the extremes. Using a declustering approach, to remove dependence, results in wider credible intervals which makes drawing conclusions about the treatment effect ambiguous. This approach also results in more negative posterior estimates of the shape parameter which results in shorter tails in the fitted GPD; thus, implying that extreme events are less likely. \\

Our findings suggest interesting possibilities for extending the modelling approach, particularly considering observed spatial dependence in the data. Incorporating methods from the spatial extremes toolbox could provide a valuable opportunity to account for this spatial dependence and potentially lead to a further increase in precision. As highlighted by \citet{arun2023leading}, while our current approach utilises stationary thresholds at the treated sites it is important to acknowledge that the LPI treatment might influence these thresholds. Therefore, exploring dynamic thresholds that capture the potential changes induced by the treatment in both the before and after periods could result in improved model fit.  Overall, our proposed approach showcases the potential for utilising extreme value Markov chain models in traffic conflict-based BA safety evaluations and our findings provide valuable insights into the effectiveness of the LPI intervention, paving the way for further investigations in this area.

\bibliography{references}
\bibliographystyle{apalike}

\newpage
\appendix
\renewcommand{\thetable}{A.\arabic{table}}
\setcounter{table}{0}
\section{Appendix}
{\scriptsize 
\begin{longtable}[c]{ccccccc}
\captionsetup{font=normalsize}
\textbf{Model}     & \multicolumn{1}{l}{Site} & \multicolumn{1}{l}{Threshold, $u$} & \multicolumn{1}{l}{} & $\beta_0$        & $\beta_1$          & $\xi$               \\ \hline
\endfirsthead
\multicolumn{7}{c}%
{{\bfseries Table \thetable\ continued from previous page}} \\
\textbf{Model}      & \multicolumn{1}{l}{Site} & \multicolumn{1}{l}{Threshold, $u$} & \multicolumn{1}{l}{} & $\beta_0$        & $\beta_1$          & $\xi$               \\ \hline
\endhead
\hline
\endfoot
\endlastfoot
\multicolumn{1}{c|}{\multirow{2}{*}{Ignore Dependence}} & \multirow{6}{*}{1}       & \multirow{6}{*}{$-5.6930$}           & Mean                 & 0.4108           & $-0.3324$            & $-0.1809$             \\
\multicolumn{1}{c|}{}                                   &                          &                                    & 95\% CI              & (0.3253, 0.4963) & ($-0.3861, -0.2786$) & ($-0.2360, -0.1259$)  \\
\multicolumn{1}{c|}{\multirow{2}{*}{Declustering}}      &                          &                                    & Mean                 & 0.9363           & $-0.2123 $           & $-0.3516$             \\
\multicolumn{1}{c|}{}                                   &                          &                                    & 95\% CI              & (0.5911, 1.2816) & ($-0.4853, 0.0607$)  & ($-0.5916, -0.1115$)  \\
\multicolumn{1}{c|}{\multirow{2}{*}{Logistic}}          &                          &                                    & Mean                 & 0.3981           & $-0.3436 $           & $-0.1649$             \\
\multicolumn{1}{c|}{}                                   &                          &                                    & 95\% CI              & (0.3015, 0.4948) & ($-0.4559, -0.2314$) & ($-0.2221, -0.1077$)  \\ \cline{2-7} 
\multicolumn{1}{c|}{\multirow{2}{*}{Ignore Dependence}} & \multirow{6}{*}{2}       & \multirow{6}{*}{$-4.8800$}           & Mean                 & 0.1058           & $-0.0019 $           & $-0.1373 $            \\
\multicolumn{1}{c|}{}                                   &                          &                                    & 95\% CI              & (0.0147, 0.1968) & ($-0.0549, 0.0512$)  & ($-$0.1949,  $-$0.0797) \\
\multicolumn{1}{c|}{\multirow{2}{*}{Declustering}}      &                          &                                    & Mean                 & 0.6136           & $-$0.0642            & $-$0.2552             \\
\multicolumn{1}{c|}{}                                   &                          &                                    & 95\% CI              & (0.2878, 0.9394) & ($-$0.3605, 0.2321)  & ($-$0.4900, $-$0.0204)  \\
\multicolumn{1}{c|}{\multirow{2}{*}{Logistic}}          &                          &                                    & Mean                 & 0.1297           & $-$0.0129            & $-$0.1454             \\
\multicolumn{1}{c|}{}                                   &                          &                                    & 95\% CI              & (0.0251, 0.2344) & ($-$0.1234, 0.0975)  & ($-$0.2082, $-$0.0826)  \\ \cline{2-7} 
\multicolumn{1}{c|}{\multirow{2}{*}{Ignore Dependence}} & \multirow{6}{*}{3}       & \multirow{6}{*}{$-$5.9500}           & Mean                 & 0.4687           & $-$0.2308            & $-$0.2119             \\
\multicolumn{1}{c|}{}                                   &                          &                                    & 95\% CI              & (0.3967, 0.5408) & ($-$0.2841, $-$0.1775) & ($-$0.2538, $-$0.1700)  \\
\multicolumn{1}{c|}{\multirow{2}{*}{Declustering}}      &                          &                                    & Mean                 & 0.9325           & $-$0.1498            & $-$0.3003             \\
\multicolumn{1}{c|}{}                                   &                          &                                    & 95\% CI              & (0.6369, 1.2280) & ($-$0.4786, 0.1790)  & ($-$0.5286, $-$0.0719)  \\
\multicolumn{1}{c|}{\multirow{2}{*}{Logistic}}          &                          &                                    & Mean                 & 0.4695           & $-$0.2033            & $-$0.1910             \\
\multicolumn{1}{c|}{}                                   &                          &                                    & 95\% CI              & (0.3858, 0.5533) & ($-$0.3139, $-$0.0926) & ($-$0.2368, $-$0.1452)  \\ \cline{2-7} 
\multicolumn{1}{c|}{\multirow{2}{*}{Ignore Dependence}} & \multirow{6}{*}{4}       & \multirow{6}{*}{$-$5.7400}           & Mean                 & 0.4632           & $-$0.3357            & $-$0.2066             \\
\multicolumn{1}{c|}{}                                   &                          &                                    & 95\% CI              & (0.3821, 0.5442) & ($-$0.3881, $-$0.2834) & ($-$0.2502, $-$0.1630)  \\
\multicolumn{1}{c|}{\multirow{2}{*}{Declustering}}      &                          &                                    & Mean                 & 0.9938           & $-$0.3250            & $-$0.3682             \\
\multicolumn{1}{c|}{}                                   &                          &                                    & 95\% CI              & (0.6649, 1.3226) & ($-$0.5854, $-$0.0647) & ($-$0.6117, $-$0.1247)  \\
\multicolumn{1}{c|}{\multirow{2}{*}{Logistic}}          &                          &                                    & Mean                 & 0.4792           & $-$0.3447            & $-$0.2071             \\
\multicolumn{1}{c|}{}                                   &                          &                                    & 95\% CI              & (0.3865, 0.5719) & ($-$0.4529, $-$0.2364) & ($-$0.2630, $-$0.1513)  \\ \cline{2-7} 
\multicolumn{1}{c|}{\multirow{2}{*}{Ignore Dependence}} & \multirow{6}{*}{5}       & \multirow{6}{*}{$-$5.9600}           & Mean                 & 0.4687           & $-$0.2308            & $-$0.2119             \\
\multicolumn{1}{c|}{}                                   &                          &                                    & 95\% CI              & (0.3967, 0.5408) & ($-$0.2841, $-$0.1775) & ($-$0.2538, $-$0.1700)  \\
\multicolumn{1}{c|}{\multirow{2}{*}{Declustering}}      &                          &                                    & Mean                 & 0.9325           & $-$0.1498            & $-$0.3003             \\
\multicolumn{1}{c|}{}                                   &                          &                                    & 95\% CI              & (0.6369, 1.2280) & ($-$0.4786, 0.1790)  & ($-$0.5286, $-$0.0719)  \\
\multicolumn{1}{c|}{\multirow{2}{*}{Logistic}}          &                          &                                    & Mean                 & 0.4695           & $-$0.2033            & $-$0.1910             \\
\multicolumn{1}{c|}{}                                   &                          &                                    & 95\% CI              & (0.3858, 0.5533) & ($-$0.3139, $-$0.0926) & ($-$0.2369, $-$0.1452)  \\ \cline{2-7} 
\multicolumn{1}{c|}{\multirow{2}{*}{Ignore Dependence}} & \multirow{6}{*}{6}       & \multirow{6}{*}{$-$6.7500}           & Mean                 & 0.409            & $-$0.2011            & $-$0.1837             \\
\multicolumn{1}{c|}{}                                   &                          &                                    & 95\% CI              & (0.3245, 0.4935) & ($-$0.2544, $-$0.1478) & ($-$0.2309, $-$0.1364)  \\
\multicolumn{1}{c|}{\multirow{2}{*}{Declustering}}      &                          &                                    & Mean                 & 0.8900           & $-$0.2082            & $-$0.2974             \\
\multicolumn{1}{c|}{}                                   &                          &                                    & 95\% CI              & (0.5393, 1.2407) & ($-$0.5297, 0.1133)  & ($-$0.4987, $-$0.0962)  \\
\multicolumn{1}{c|}{\multirow{2}{*}{Logistic}}          &                          &                                    & Mean                 & 0.4177           & $-$0.2151            & $-$0.1719             \\
\multicolumn{1}{c|}{}                                   &                          &                                    & 95\% CI              & (0.3211, 0.5144) & ($-$0.3240, $-$0.1062) & ($-$0.2229, $-$0.1208)  \\ \cline{2-7} 
\multicolumn{1}{c|}{\multirow{2}{*}{Ignore Dependence}} & \multirow{6}{*}{7}       & \multirow{6}{*}{$-$5.5300}           & Mean                 & 0.3845           & $-$0.1920            & $-$0.1823             \\
\multicolumn{1}{c|}{}                                   &                          &                                    & 95\% CI              & (0.3051, 0.4639) & ($-$0.2411, $-$0.1428) & ($-$0.2256, $-$0.1390)  \\
\multicolumn{1}{c|}{\multirow{2}{*}{Declustering}}      &                          &                                    & Mean                 & 0.9984           & $-$0.1637            & $-$0.4098             \\
\multicolumn{1}{c|}{}                                   &                          &                                    & 95\% CI              & (0.6883, 1.3084) & ($-$0.4064,  0.0791) & ($-$0.6374, $-$0.1822)  \\
\multicolumn{1}{c|}{\multirow{2}{*}{Logistic}}          &                          &                                    & Mean                 & 0.3999           & $-$0.1763            & $-$0.2046             \\
\multicolumn{1}{c|}{}                                   &                          &                                    & 95\% CI              & (0.3082, 0.4915) & ($-$0.2786, $-$0.0740) & ($-$0.2541, $-$0.1551)  \\ \cline{2-7} 
\multicolumn{1}{c|}{\multirow{2}{*}{Ignore Dependence}} & \multirow{6}{*}{8}       & \multirow{6}{*}{$-$7.4020}           & Mean                 & 0.3183           & $-$0.0925            & $-$0.0834             \\
\multicolumn{1}{c|}{}                                   &                          &                                    & 95\% CI              & (0.2354, 0.4012) & ($-$0.1523, $-$0.0327) & ($-$0.1317, $-$0.0351)  \\
\multicolumn{1}{c|}{\multirow{2}{*}{Declustering}}      &                          &                                    & Mean                 & 1.0146           & $-$0.1652            & $-$0.2931             \\
\multicolumn{1}{c|}{}                                   &                          &                                    & 95\% CI              & (0.6750, 1.3542) & ($-$0.4812, 0.1508)  & ($-$0.5060, $-$0.0801)  \\
\multicolumn{1}{c|}{\multirow{2}{*}{Logistic}}          &                          &                                    & Mean                 & 0.3153           & $-$0.0895            & $-$0.1186             \\
\multicolumn{1}{c|}{}                                   &                          &                                    & 95\% CI              & (0.2190, 0.4116) & ($-$0.2136,  0.0345) & ($-$0.1728, $-$0.0643)  \\ \cline{2-7} 
\multicolumn{1}{c|}{\multirow{2}{*}{Ignore Dependence}} & \multirow{6}{*}{9}       & \multirow{6}{*}{$-$6.1300}           & Mean                 & 0.1779           & 0.0295             & $-$0.1706             \\
\multicolumn{1}{c|}{}                                   &                          &                                    & 95\% CI              & (0.0718, 0.2841) & ($-$0.0366,  0.0955) & ($-$0.2325, $-$0.1088)  \\
\multicolumn{1}{c|}{\multirow{2}{*}{Declustering}}      &                          &                                    & Mean                 & 0.8241           & 0.2195             & $-$0.4150             \\
\multicolumn{1}{c|}{}                                   &                          &                                    & 95\% CI              & (0.4972, 1.1510) & ($-$0.0773, 0.5163)  & ($-$0.6876, $-$0.1424)  \\
\multicolumn{1}{c|}{\multirow{2}{*}{Logistic}}          &                          &                                    & Mean                 & 0.2068           & 0.0404             & $-$0.1358             \\
\multicolumn{1}{c|}{}                                   &                          &                                    & 95\% CI              & (0.0859, 0.3276) & ($-$0.0971,  0.1779) & ($-$0.2066, $-$0.0650)  \\ \cline{2-7} 
\multicolumn{1}{c|}{\multirow{2}{*}{Ignore Dependence}} & \multirow{6}{*}{10}      & \multirow{6}{*}{$-$5.1600}           & Mean                 & 0.1200           & $-$0.0289            & $-$0.1340             \\
\multicolumn{1}{c|}{}                                   &                          &                                    & 95\% CI              & (0.0279, 0.2122) & ($-$0.0874,  0.0296) & ($-$0.1901, $-$0.0778)  \\
\multicolumn{1}{c|}{\multirow{2}{*}{Declustering}}      &                          &                                    & Mean                 & 0.6670           & $-$0.0939            & $-$0.2505             \\
\multicolumn{1}{c|}{}                                   &                          &                                    & 95\% CI              & (0.3498, 0.9843) & ($-$0.4052, 0.2175)  & ($-$0.4641, $-$0.0369)  \\
\multicolumn{1}{c|}{\multirow{2}{*}{Logistic}}          &                          &                                    & Mean                 & 0.1327           & $-$0.0303            & $-$0.1072             \\
\multicolumn{1}{c|}{}                                   &                          &                                    & 95\% CI              & (0.0286, 0.2368) & ($-$0.1500, 0.0894)  & ($-$0.1741, $-$0.0402)  \\ \cline{2-7} 
\multicolumn{1}{c|}{\multirow{2}{*}{Ignore Dependence}} & \multirow{6}{*}{11}      & \multirow{6}{*}{$-$6.5155}           & Mean                 & 0.2028           & $-$0.0541            & $-$0.1249             \\
\multicolumn{1}{c|}{}                                   &                          &                                    & 95\% CI              & (0.1095, 0.2960) & ($-$0.1124,  0.0042) & ($-$0.1661, $-$0.0837)  \\
\multicolumn{1}{c|}{\multirow{2}{*}{Declustering}}      &                          &                                    & Mean                 & 0.7097           & $-$0.2032            & $-$0.1831             \\
\multicolumn{1}{c|}{}                                   &                          &                                    & 95\% CI              & (0.3112, 1.1082) & ($-$0.5839,  0.1776) & ($-$0.4316,  0.0654)  \\
\multicolumn{1}{c|}{\multirow{2}{*}{Logistic}}          &                          &                                    & Mean                 & 0.2102           & $-$0.0701            & $-$0.1239             \\
\multicolumn{1}{c|}{}                                   &                          &                                    & 95\% CI              & (0.1033, 0.3171) & ($-$0.1910, 0.0508)  & ($-$0.1809, $-$0.0668)  \\ \cline{2-7} 
\multicolumn{1}{c|}{\multirow{2}{*}{Ignore Dependence}} & \multirow{6}{*}{12}      & \multirow{6}{*}{$-$6.2600}           & Mean                 & 0.3638           & $-$0.0929            & $-$0.1961             \\
\multicolumn{1}{c|}{}                                   &                          &                                    & 95\% CI              & (0.2757, 0.4518) & ($-$0.1476, $-$0.0382) & ($-$0.2467, $-$0.1455)  \\
\multicolumn{1}{c|}{\multirow{2}{*}{Declustering}}      &                          &                                    & Mean                 & 1.0513           & $-$0.2451            & $-$0.3788             \\
\multicolumn{1}{c|}{}                                   &                          &                                    & 95\% CI              & (0.6634, 1.4393) & ($-$0.5202, 0.0300)  & ($-$0.6586, $-$0.0989)  \\
\multicolumn{1}{c|}{\multirow{2}{*}{Logistic}}          &                          &                                    & Mean                 & 0.3778           & $-$0.0941            & $-$0.1688             \\
\multicolumn{1}{c|}{}                                   &                          &                                    & 95\% CI              & (0.2774, 0.4783) & ($-$0.2086, 0.0205)  & ($-$0.2255, $-$0.1121)  \\ \cline{2-7} 
\multicolumn{1}{c|}{\multirow{2}{*}{Ignore Dependence}} & \multirow{6}{*}{13}      & \multirow{6}{*}{$-$5.0500}           & Mean                 & 0.2562           & $-$0.1708            & $-$0.1470             \\
\multicolumn{1}{c|}{}                                   &                          &                                    & 95\% CI              & (0.1641, 0.3482) & ($-$0.2259, $-$0.1157) & ($-$0.1919, $-$0.1021)  \\
\multicolumn{1}{c|}{\multirow{2}{*}{Declustering}}      &                          &                                    & Mean                 & 0.9524           & $-$0.3738            & $-$0.3970             \\
\multicolumn{1}{c|}{}                                   &                          &                                    & 95\% CI              & (0.6237, 1.2811) & ($-$0.6159, $-$0.1318) & ($-$0.6729, $-$0.1212)  \\
\multicolumn{1}{c|}{\multirow{2}{*}{Logistic}}          &                          &                                    & Mean                 & 0.2456           & $-$0.1858            & $-$0.1634             \\
\multicolumn{1}{c|}{}                                   &                          &                                    & 95\% CI              & (0.1394, 0.3518) & ($-$0.2982, $-$0.0734) & ($-$0.2260, $-$0.1008)  \\ \cline{2-7} 
\multicolumn{1}{c|}{\multirow{2}{*}{Ignore Dependence}} & \multirow{6}{*}{14}      & \multirow{6}{*}{$-$10.4100}          & Mean                 & 0.3635           & $-$0.1695            & $-$0.0788             \\
\multicolumn{1}{c|}{}                                   &                          &                                    & 95\% CI              & (0.2895, 0.4374) & ($-$0.2394, $-$0.0996) & ($-$0.1207, $-$0.0368)  \\
\multicolumn{1}{c|}{\multirow{2}{*}{Declustering}}      &                          &                                    & Mean                 & 0.7947           & $-$0.1247            & $-$0.0480             \\
\multicolumn{1}{c|}{}                                   &                          &                                    & 95\% CI              & (0.4865, 1.1029) & ($-$0.5910, 0.3417)  & ($-$0.2650, 0.1689)   \\
\multicolumn{1}{c|}{\multirow{2}{*}{Logistic}}          &                          &                                    & Mean                 & 0.3666           & $-$0.1649            & $-$0.0572             \\
\multicolumn{1}{c|}{}                                   &                          &                                    & 95\% CI              & (0.2836, 0.4496) & ($-$0.3082, $-$0.0216) & ($-$0.1056 $-$0.0087    \\ \cline{2-7} 
\multicolumn{1}{c|}{\multirow{2}{*}{Ignore Dependence}} & \multirow{6}{*}{15}      & \multirow{6}{*}{$-$5.1635}           & Mean                 & 0.2499           & $-$0.0947            & $-$0.1548             \\
\multicolumn{1}{c|}{}                                   &                          &                                    & 95\% CI              & (0.1612, 0.3386) & ($-$0.1514, $-$0.0380) & ($-$0.2156, $-$0.0940)  \\
\multicolumn{1}{c|}{\multirow{2}{*}{Declustering}}      &                          &                                    & Mean                 & 0.6906           & $-$0.1064            & $-$0.2655             \\
\multicolumn{1}{c|}{}                                   &                          &                                    & 95\% CI              & (0.3828, 0.9985) & ($-$0.4114,  0.1986) & ($-$0.4958, $-$0.0352)  \\
\multicolumn{1}{c|}{\multirow{2}{*}{Logistic}}          &                          &                                    & Mean                 & 0.2289           & $-$0.0901            & $-$0.1454             \\
\multicolumn{1}{c|}{}                                   &                          &                                    & 95\% CI              & (0.1274, 0.3304) & ($-$0.2066, 0.0264)  & ($-$0.2074, $-$0.0833)  \\ \hline
\caption{\normalsize{Numerical posterior summaries for model parameters at all intersections. For each parameter we show the posterior mean and 95\% credible interval from the three analyses (ignoring dependence, declustering and logistic) and the chosen threshold.} }
\label{tab:FullResults}
\end{longtable}
}
\end{document}